\documentclass[aps,twocolumn,showpacs,english]{revtex4}

\usepackage[T1]{fontenc}
\usepackage[latin9]{inputenc}
\usepackage{textcomp}
\usepackage{amsmath}
\usepackage{graphicx}
\usepackage{amssymb}
\usepackage{babel}

\begin{document}

\title{Photon correlation spectroscopy on a single  quantum dot embedded in a  nanowire}

\author{G. Sallen$^{1}$, A. Tribu$^{2}$, T. Aichele$^{1}$, R. Andr\'{e}$^{1}$, C. Bougerol$^{1}$, S. Tatarenko$^{1}$, K. Kheng$^{2}$, and J. Ph.~Poizat$^{1}$}

\affiliation{CEA-CNRS-UJF group 'Nanophysique et Semiconducteurs',\\
$^1$ Institut N\'{e}el, CNRS - Universit\'{e} Joseph Fourier, 38042 Grenoble, France, \\
$^2$ CEA/INAC/SP2M, 38054 Grenoble, France}

\begin{abstract}
We have observed strong photoluminescence from a single CdSe quantum dot embedded in a ZnSe nanowire. Exciton, biexciton and charged exciton lines have been identified unambiguously using photon correlation spectroscopy. This technique has provided a detailed picture of the dynamics of this new system.
 This type of semi conducting quantum dot turns out to be a very efficient single photon source in the visible.
Its particular growth technique opens new possibilities as compared to the usual self-asssembled quantum dots.

\end{abstract}


\maketitle

Semiconductor nanowires (NWs) appear as  promising building blocks for nanoscale devices and circuits with  impressive potential applications including nanoelectronics \cite{Duan01,Lu,The03}, optoelectronics (light emitting diodes  \cite{Konenkamp,Kim}, nanolasers \cite{Duan03}),  thermoelectrical energy conversion  \cite{Hochbaum}, and biological or chemical sensors \cite{Cui}.
Moreover, high quality defect free nanowires can be grown on low-cost, routinely used substrates such as silicon, which means that they could easily be used for fabricating commercial devices and  could possibly be integrated with mainstream Si microelectronics devices.

NW growth methods allow for the variation of the chemical composition \cite{Gud02,Bjork} or doping \cite{Yang} along the longitudinal or radial directions. This enables the fabrication of well controlled  1D nanoscale  heterostructures \cite{Bjork}. For example, as shown in this work, it is possible to insert a slice of a low band gap semiconductor within a high bandgap NW and thus realize a light emitting quantum dot (QD) operating as a single photon source \cite{borgstrom,Tribu}. So far, work on the light emitting properties of  single quantum dots has mainly concerned self-assembled QDs  formed by surface forces induced by lattice mismatch between different materials. Such QDs have been widely used in the past decade as single photon sources \cite{Michler} and for their potential application in quantum information processing (see for example \cite{Fushman,Hanson}).
QDs in NWs appear to be an interesting alternative to self-assembled quantum dots. The absence of a wetting layer offers a better confinement which  could enable room temperature production of
 single photons \cite{Tribu}.
Radial growth techniques enable engineering of optical guides allowing more efficient light extraction than in bulk materials \cite{Pauzauskie,Gregersen}.
Furthermore,  NW based heterostructures, being much less limited by lattice mismatches, greatly widen the possible materials combinations and
enable well controlled stacking of several QDs in a single NW,   offering interesting possibilities for quantum information processing \cite{Simon}.

 In this letter we present the first detailed optical characterization of excitonic emission in a single CdSe QD embedded in a ZnSe NW.
 We have already shown that this system is an efficient single photon source operating at temperature as high as 220 K \cite{Tribu}.
 Single photon emission from NWs has otherwise only been demonstrated at  4 K in InAs QDs embedded in InP NWs \cite{borgstrom}.
 Our system emits light around $550$ nm ($2.2$eV) where silicon avalanche photodiodes (APDs) are very efficient.
 This has allowed us to
 perform the first thorough  spectroscopic analysis   of a QD embedded in a NW by using photon correlation spectroscopy \cite{Kiraz}. We have  identified unambiguously the exciton, biexciton, and charged exciton lines and obtained information on the charging dynamics of this QD.

\begin{figure}
 \resizebox{0.3\textwidth}{!}{\includegraphics{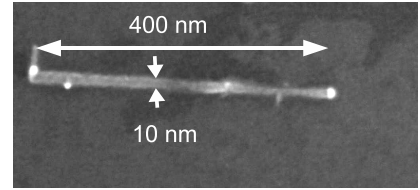}}
 \caption{  Scanning electron microscope image of a single CdSe/ZnSe nanowire deposited on a silicon substrate.   }
 \label{fig:imageSEM}
\end{figure}

The wires are grown by Molecular Beam Epitaxy (MBE) in the Vapour-Liquid-Solid (VLS) growth
mode catalysed by gold particles on a Si substrate.
In order to make QDs, a small region of CdSe is inserted in  the ZnSe NW.
This is done by interrupting the ZnSe growth,
changing to CdSe growth for a short time and then growing  ZnSe again \cite{Tribu}.
Details about the growth of the ZnSe NWs can be found in reference \cite{Aichele}.
The wire diameter (around $10$ nm) is of
the order of the bulk exciton Bohr diameter for CdSe (11
nm). This means that the carriers in the CdSe QD are
in the strong quantum confinement regime. For the study of single
NWs, the sample is sonicated in methanol, causing  NWs to break off the substrate into
the solution.  Droplets of this
solution are then deposited on a Si substrate, and a
 low density of individual NWs is obtained after evaporation.
 As shown in fig. \ref{fig:imageSEM}, individual NWs can be isolated, allowing single QD optical spectroscopy.

The experimental apparatus is a standard  microphotoluminescence ($\mu$PL) set-up. The
samples are mounted on a XYZ piezo motor system in a He flow cryostat  at a temperature of
4 K. The optical excitation is provided by a 405 nm continuous-wave (CW) diode laser illuminating the sample
via a microscope objective of numerical aperture $NA=0.65$ located in the cryostat.
The NW emission is
 collected by the same objective  and
 sent  to a 50/50
beamsplitter for correlation measurements. In each
arm of the beamsplitter, the light is dispersed by a monochromator
(1200 grooves/mm grating, 30 cm and 50 cm focal length
respectively). Each monochromator has a switchable mirror inside,
which can direct the luminescence either onto a charge coupled device (CCD) camera for the
measurement of the PL spectrum or through the exit slit towards
a low jitter (40 ps), high quantum efficiency APD.
The detectors send electrical pulses into
a time-correlated single photon module that
builts an histogram of the time delays between successive photons.
This histogram is proportional to the second order correlation function $g^{(2)}(t)$
 \cite{Michler}.
 The overall
temporal resolution of our set-up is essentially limited by the
jitter of the APDs and the dispersion of the monochromator
gratings. This time resolution was measured by recording the
autocorrelation function of 1 ps  pulses from a
frequency-doubled Ti:Sapphire laser. A full width at half maximum
 of 90 ps was obtained for the autocorrelation function
peak.

\begin{figure}
 \resizebox{0.45\textwidth}{!}{\includegraphics{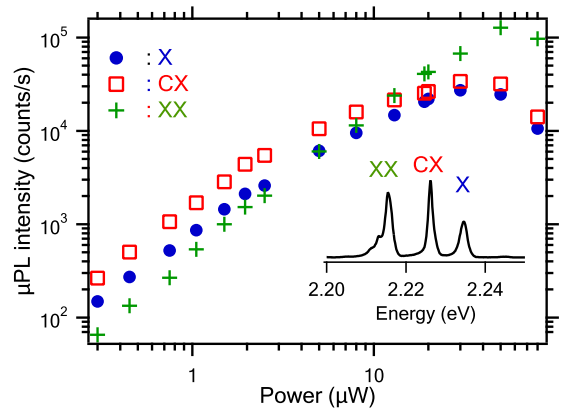}}
 \caption{  Inset : $\mu$PL spectrum at an excitation power of $P=  15 µW $. The electronic background noise has been subtracted. Main plot : Line intensities as a function of excitation power. }
 \label{fig:spectra}
\end{figure}

The results presented in this work  all come from the same QD.
A typical $\mu$PL spectrum is shown in the inset of fig. \ref{fig:spectra} where three lines can be seen.
A comparison with relative energy positions of known emission lines in spectra of self-assembled CdSe/ZnSe QDs \cite{Turck,Patton} suggests that these lines
 correspond to the exciton (X), the biexciton (XX) and the charged exciton (CX).
Unambiguous proof for the assignement of these lines will be given below using photon correlation spectroscopy.
 The width of the lines is due to spectral diffusion \cite{these}.

One of the characteristic features of such  NW QD structures
is their polarization properties. As shown in \cite{Tribu}
 the excitation efficiency and the luminescence are
both strongly polarization dependent \cite{Wang, Niquet,vanWeert}.
A $90\%$ contrast is obtained for the  excitation efficiency depending on the direction of the  linear polarization of the pumping laser.
The light emission is also $90\%$ linearly polarized in the same direction as the best pumping polarization,  independently of the excitation polarization.

It can be seen in fig. \ref{fig:spectra} that  the saturation level of the XX line is more than three times larger than that of the X line. This is due to a strong storage effect of the dark exciton state owing to a rather  large dark-bright exciton splitting ($\Delta E= 6$ meV) \cite{Dark}. The dark exciton state reduces the intensity of the X line owing to the leakage from the bright to the dark exciton states caused by spin flip but it remains an efficient intermediate state for populating the XX state \cite{Reischle}.

\begin{figure}
 \resizebox{0.45\textwidth}{!}{\includegraphics{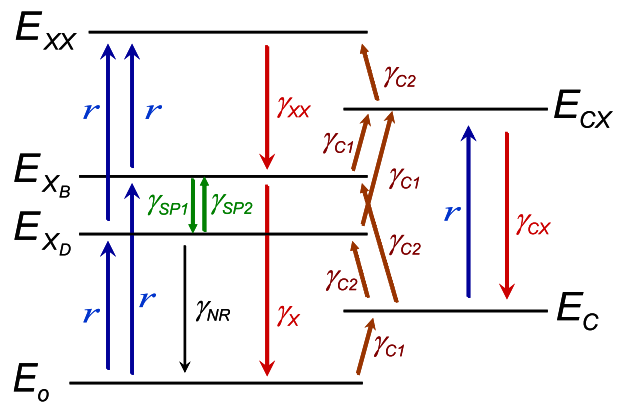}}
 \caption{Level scheme including the empty dot ($E_0$), the dark exciton ($E_{X_D}$), the bright exciton ($E_{X_B}$), the biexciton ($E_{XX}$), the charged dot ($E_{C}$), and the charged exciton ($E_{CX}$). The various rates between the different level are indicated in the figure.}
 \label{fig:model}
\end{figure}

The level scheme  used to model our system is shown in figure \ref{fig:model}.
It is based on models used in references \cite{Santori,Baier,Suffczynski} where
the carriers can enter the QD either individually (power dependent rates $\gamma_{C1}$ and $\gamma_{C2}$) or already bound as excitons (power dependent rate $r$).
We have added the dark exciton, which plays a key role in our system.
It includes the bright and dark exciton,  the charged exciton, and the biexciton. A triexciton level and a charged biexciton level are also included to avoid artificial saturation of the biexciton and of the charged exciton but they are not represented in figure \ref{fig:model}.

All of the power independent parameters of the model can be evaluated independently prior to photon correlation experiments by performing lifetime measurements \cite{Dark}.
Their values  are listed in table \ref{table:parameters}. These values are compatible with what has already been measured on self-assembled CdSe/ZnSe QDs   \cite{Patton}.

\begin{table}
\label{tab:resumfix}
\begin{tabular}{|c|c|c|c|c|c|c|}
\hline
$\gamma_X$ & $\gamma_{XX}$  & $\gamma_{CX}$  & $\gamma_{SP_1}$ & $\gamma_{SP_2}$ & $\gamma_{NR}$ \\
\hline 1.4&2.5&1.7&1.4&0&0.2\\
\hline
\end{tabular}
 \caption{Table of power independent transition rates in $ns^{-1}$. }
 \label{table:parameters}
\end{table}

\begin{figure}
 \resizebox{0.45\textwidth}{!}{\includegraphics{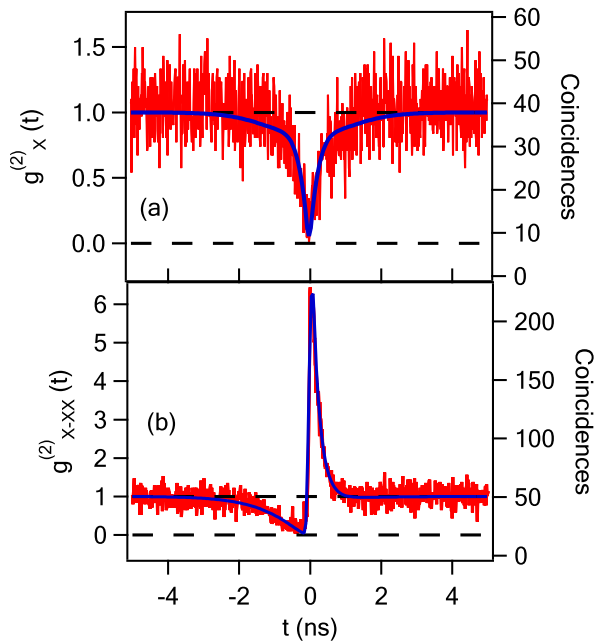}}
 \caption{ (a) Exciton emission autocorrelation,  (b) exciton-biexciton cross-correlation
 for an excitation power $P=15 \mu W$.
  The left axes are the correlation functions
corrected from the background and the right axes are the raw coincidence
rates (see text).
The fit is performed using the model based on fig. \ref{fig:model}. The power dependent parameters used for the fit are
$r=0.6 ns^{-1}$, $\gamma_{C1}=\gamma_{C2}=1 ns^{-1}$. The other parameters are given in table \ref{table:parameters}.}
 \label{fig:X-XX}
\end{figure}

We come now to the main results of this work on photon correlation experiments. We present first the data concerning the neutral QD in fig. \ref{fig:X-XX}. The autocorrelation of the X line emission is shown in fig. \ref{fig:X-XX} (a) exhibiting a clear antibunching which is characteristic of the  statistics
of a single photon emitter \cite{Michler}.

Fig \ref{fig:X-XX} (b) shows the cross correlation measurement between
the X and the XX line. It displays the typical asymmetric shape with bunching and antibunching
 features that is the signature for the cascaded emission of a XX photon followed by a X photon \cite{Moreau}. This allows us to identify unambiguously these two lines as  exciton and biexciton of the same QD.
 Note that the narrow bunching peak can only be fitted  if the dark exciton is included in the model.

\begin{figure}
 \resizebox{0.45\textwidth}{!}{\includegraphics{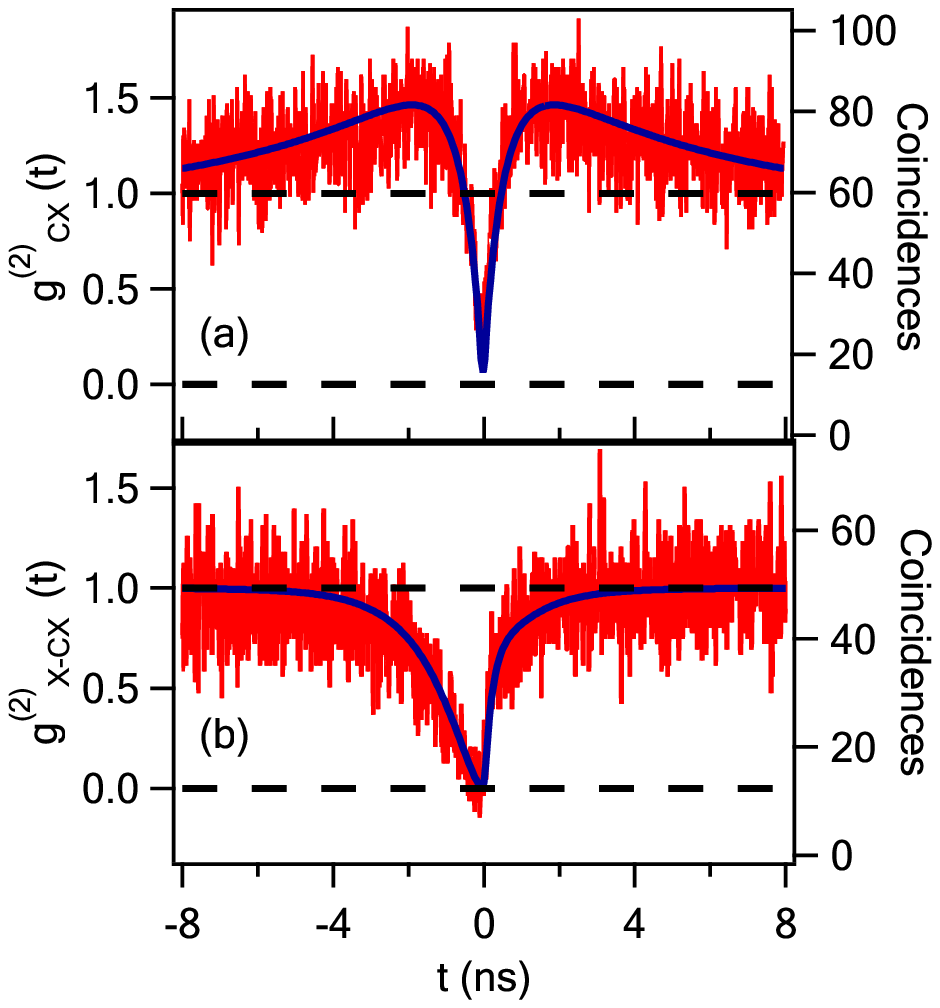}}
 \resizebox{0.3\textwidth}{!}{\includegraphics{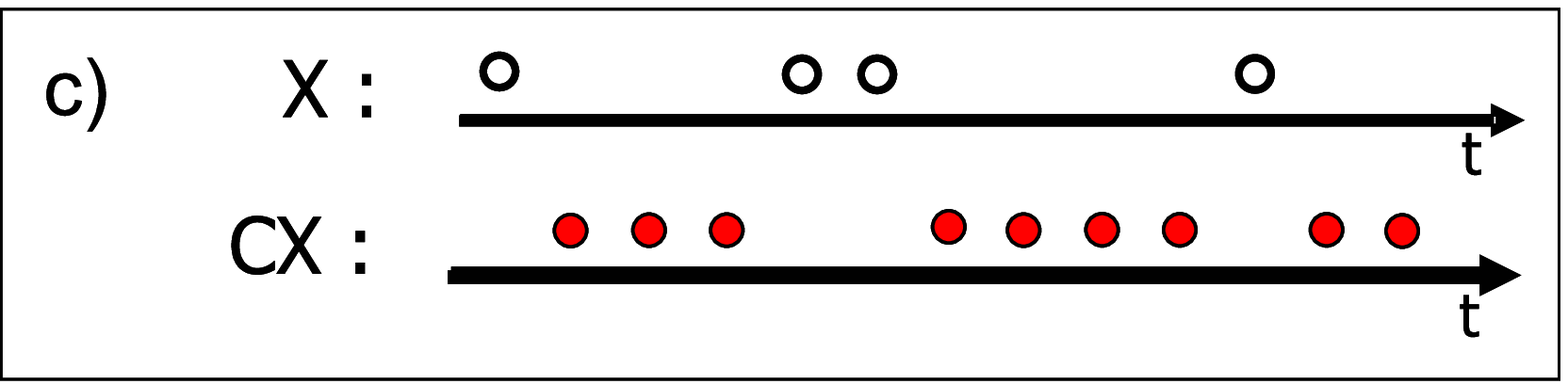}}
 \caption{(a) Charged exciton autocorrelation for an excitation power $P= 8 \mu W$, (b) Exciton-Charged exciton cross-correlation for an excitation power $P= 10 \mu W$.
 The fits are performed using the model based on fig. \ref{fig:model}. The power dependent parameters used for the fits  are respectively
$r=0.31 ns^{-1}$, $\gamma_{C1}= 0.09 ns^{-1}$, $\gamma_{C2}=0.058 ns^{-1}$ for the CX autocorrelation (a) and $r=0.37 ns^{-1}$, $\gamma_{C1}= 0.25 ns^{-1}$, $\gamma_{C2}=0.28 ns^{-1}$ for the X-CX cross-correlation (b). The other parameters are given in table \ref{table:parameters}.
(c) Representation of the  stream of photons coming alternatively from the neutral and the charged QD. }
 \label{fig:X-CX}
\end{figure}

For all the correlation  graphs (figs. \ref{fig:X-XX} and \ref{fig:X-CX}) the right vertical axes are the raw number of coincidences. The left axes represent the normalized correlation function according to a Poisson statistics where the coincidences involving background photons have been subtracted. The corrected correlation function $g^{(2)}$ is related to the uncorrected one $g^{(2)}_u$ by $g^{(2)} -1= (g^{(2)}_u -1)/\rho^2 $, where $\rho=S/(S+B)$ with $S$ and $B$  respectively the number of signal and background photons as  measured in the spectrum of fig. \ref{fig:spectra}.

The autocorrelation of the CX line is shown in fig. \ref{fig:X-CX} (a). As for the X line, it exhibits a clear antibunching. Also, on a larger time scale, a bunching effect can  be observed. This is due to the hopping from the charged state to the neutral state of the QD as confirmed by the X-CX cross correlation displayed in  fig. \ref{fig:X-CX} (b). This proves that the CX line comes from the same QD as the X line. The situation is
 depicted schematically in fig. \ref{fig:X-CX} (c), which shows that the photons are
emitted either from the charged or from the neutral state of the QD. The
average time spent  by the quantum dot in the charged state is given
by the characteristic time of an exponential fitting of the bunching peak in fig. \ref{fig:X-CX} (a) which is $5$ ns.
It should be noticed that the antibunching dip in figure \ref{fig:X-CX} (b) is not symmetrical. Negative (positive) time corresponds to the probability of detecting a CX (X) photon  after having detected a X (CX) photon. Formation of a CX in an empty QD ($t<0$) requires the loading of three charges, whereas the formation of an X in a QD with a single charge (level $E_c$ in fig. \ref{fig:model}) is faster since it requires only the loading of a single charge ($t>0$) \cite{Santori,Baier,Suffczynski}.
Note that we do not know whether this charge is positive or negative. We can only say that by comparison with spectra for self-assembled CdSe/ZnSe QDs \cite{Turck,Patton} the charge state is probably negative.

As can be seen in figs. \ref{fig:X-XX} and \ref{fig:X-CX} the experimental results are  fitted very well by the  model shown in fig.\ref{fig:model} taking into account the temporal resolution of our experimental set-up ($90$ ps). Inclusion of the dark exciton
turned out to be essential for the modelling of the photon correlation data.
Allowing for the coexistence of two excitation mechanisms, namely charge by charge (described by $\gamma_{C1}$ and $\gamma_{C2}$) or directly feeding the QD with an already
bound exciton (described by $r$) has also turned out to be necessary for the fitting.
The QD charge hopping time depends on the value of these parameters. Note that XX autocorrelation and the CX-XX cross-correlation (not shown) are also  fitted well using the same set of parameters \cite{these}. The model gives also the correct  intensities of the spectral lines within $10 \%$ \cite{these}.

In conclusion, we have used photon correlation spectroscopy to characterize a new type of light emitting quantum dot embedded in a nanowire. We obtained a very good fit to the experimental data with a model based on a standard excitonic level scheme. This  allowed us to extract quite complete dynamics of the neutral and charged excitons including charge hopping between these two states of the QD.
CdSe/ZnSe QDs in NWs are nano-objects situated between self-assembled QDs and CdSe based colloidal nanocrystals \cite{Michler2,Mahler}.
The latter operate at room temperature but have a blinking problem and a lifetime above $20$ ns.
On the other hand, self-assembled QDs are non-blinking and feature a sub-nanosecond lifetime allowing GHz  repetition rates. QDs embedded in NWs have the potential to combine  the best of both worlds by offering non-blinking room temperature \cite{Tribu} single photon sources with a high repetition rate.
Furthermore, the
versatility of this particular nanostructure
 growth technique  offers interesting perspectives for engineering semi-conducting QDs, such as coupled QDs or waveguide coupled QDs.

We thank L. Besombes for many fruitful discussions, F. Donatini for very efficient technical support, and R. Cox for a careful reading of the manuscript.
T.A. acknowledges support by Deutscher
Akademischer Austauschdienst (DAAD). Part of this
work was supported by European project QAP (Contract No. 15848).

\end{document}